# An approach for modeling of multiphase flows as random processes

IVAN V. KAZACHKOV[1,2]

[1]Dept of Information Technologies and Data Analysis,
Nizhyn Gogol State University, UKRAINE, http://www.ndu.edu.ua
[2]Dept of Energy Technology, Royal Institute of Technology, Stockholm, 10044, SWEDEN,
Ivan.Kazachkov@energy.kth.se, http://www.kth.se/itm/inst?l=en_UK

**Abstract**

The basic system of differential equations for a multiphase flow with the introduction of the probability of each phase in the flow is considered. The main analysis is focused on the case of a heterogeneous two-phase flow. The conservation equations for mass, momentum and energy are obtained under the assumption that parameters of the interacting phases are players of the statistical process. In parallel, dynamical system by the Kolmogorov's theorem for two states of a statistical system (phases of a two-phase mixture) is considered. Probability of phases in a flow is taken further for comparison with the probability and parameters of a two-phase flow from the equations of flow dynamics. Analysis of the parameters of a two-phase flow is performed as relating to available flow regimes from a statistical point of view on the basis of achievable parameter values and, first of all, on the condition that the probability is strictly in the range from 0 to 1. Correspondence of parameters by the equation array for flow dynamics and by solution of the dynamical system of two phases (two interacting statistical states) revealed the values of the coefficients for dynamical system, expressed in terms of the flow parameters. The results obtained are intended for further discussion, research, comparison with experimental data and with results of other researchers of the multiphase flows.

## 1. Physical-mathematical models of continua and modelling of multiphase flows

Creation of physical and mathematical models of continua and subsequent research of the physic-mechanical and other processes running in them is most often based on the hypotheses of a continuity of medium and continuous *n*-times differentiability of all functions describing parameters of continua, almost everywhere except for separate points, lines or surfaces on which gaps are allowed. It allows using phenomenological approach applying the developed methods of the mathematical analysis and mathematical physics. However, though phenomenological approach allowed solving a set of the problems of continuum mechanics, which became already classical and is nowadays one of the most often applied, it is necessary to account that many processes in continua don't satisfy this physical model. For example, in turbulent flow the velocity acceleration fields aren't described in a class of continuous or nearly everywhere continuous functions, and the film flow compressed in the direction of a tangent to a free surface becomes nowhere differentiable in this direction even if remains continuous (a surface saw-tooth with teeth, perpendicular to it). The other examples are: spraying and cavitation. Here it is impossible to carry out individualization even volumes since the continual medium turns into a set of free points.

In heterogeneous media the fields of velocities, temperatures, etc. are fractured and the combination of two various fields in one continuum, a polysemy of parameters belong not to the individualized point of a medium but to a space point in which the individualized points of the various parameters are combined in. Statistical approach and various options of variation methods [1-6] are applied to the non-classical problems of the continuum mechanics, which aren't satisfying to hypotheses of the phenomenological theory. In view of mathematical complexity statistical (microscopic) approach is most often used for justification of phenomenological (macroscopic) models of continua if only it isn't unique: the discharged gases, plasma, etc. Strictly speaking, in nature there are no real continua, however, the continuity hypothesis describes them well at the macro level and the model of continua allows using the powerful theory of continuous functions, the differential and integral calculus.

For the systems, which do not answer continuity hypothesis of an occupied space, the fractal [7] theory (objects of fractional dimension) and the developed integral and differential calculus of arbitrary order [8, 9] (not only integer derivations and integrals as in classical calculus) are more adequate.  Here we use phenomenological approach at which creation of mathematical models of continua is based on an assumption that each point of the medium (physically infinitesimal volume), which physical and mathematical state is characterized by a set of the defining parameters introduced on the basis of experimental data and theoretical investigations. Now from a position of phenomenological approach, such set of mathematical models of various classes of tasks taking into account their specific features have been developed that a need ripened for their systematization and development the basic principles of mathematical modeling of processes in continua. Therefore, when developing new complexes of mathematical and numerical models it is necessary to proceed from the modular principle allowing to unify as much as the possible process of modeling and to facilitate the use of mathematical and computer numerical models by various researchers in various tasks.

At the phenomenological approach, a creation of mathematical models of continua is based on an assumption that each point of the medium (physically infinitesimal volume), which physic-mechanical state is characterized by a set of the defining parameters introduced on the basis of experimental and theoretical data or statistically averaged functions (temperature, for example). The general equations of dynamics of continua by any structure (including the heterogeneous mix considered without phase interaction, which can be not taken into account when studying the movement of the heterogeneous system as a uniform complex continuous medium), may be represented in a form:

$$\partial \rho / \partial t = -div(\rho \vec{v}), \quad (1)$$

$$\rho(\partial \vec{v}/\partial t + \vec{v}\nabla\vec{v}) = divP + \rho\vec{F} - \sum_{j=1}^{N} \vec{v}\nabla(\rho_j \vec{v}_j), \quad (2)$$

$$\rho(\partial e/\partial t + \vec{v}\nabla e) = div(\vec{q} + P\vec{v}) + \rho\vec{F}\vec{v} + \sum_{j=1}^{N}\left[\rho_j\vec{F}_j\vec{v} - div(\rho_j e_j \vec{v})\right], \quad (3)$$

where $\rho$ - the density of the heterogeneous medium, $\vec{v}$ - velocity vector for heterogeneous medium, $t$- time, $P$- stress tensor, $\vec{F}$ - volumetric force, $\rho_j, \vec{v}_j$ - parameters of the medium's components (similar for the other parameters with indexes), $e$- the specific density of energy, $q$ - specific volumetric energy influx.

The first equation (1) expresses mass conservation law, the second (2) - the balance of an impulse, the third (3) – energy conservation. For reversible processes, the uncompensated warmth is equal to zero. Except for internal energy and entropy the other functions of state and additional thermodynamic relations are used. In case of the heterogeneous medium when an exchange of mass, impulse, and energy between phases inside the volume or at the boundaries must be taken into account, the terms on the exchange of mass, impulse and energy between phases of the heterogeneous mix in the equation array (1) - (3) must be explicitly specified. Namely, this makes the main problem in mechanics of heterogeneous media since it is in most cases not clear how to define corresponding intensities of the mass, impulse and energy exchange between the phases of heterogeneous continua.

## 2. Statistical approach for modelling of the multiphase flows

Let us consider that each phase is present in given point of continuum medium with a probability Let us consider the interaction of the phases of multiphase flow according to the theorem of Kolmogorov [10, 11] for the Markov processes. Locally, the interaction of the phases can be considered, and the Markov processes of two states (interacting phases 1 and 2) or even more (in each point of the medium a few phases in a flow may interact simultaneously) can be analyzed (see in Fig. 1 schematically).

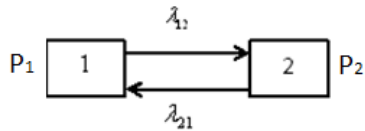

Fig. 1 Interacting phases as the Markov process

Here $P_i$ is the probability of the $i$-phase at a given point of the medium, $\lambda_{ij}$ is the intensity of the transition of the state $i$ to the state $j$. If a few phases are interacting, then all the phases are represented as shown in Fig. 1 as interacting actors of the Markov process. The intensities of the phases' interaction are determined by physical properties and flow regime. According to the Kolmogorov theorem these probabilities are calculated from the following equations

$$\frac{dP_i}{dt} = -\sum_{j=1}^{m} \lambda_{ij} P_i + \sum_{j=1}^{m} \lambda_{ji} P_j, \qquad \sum_{i=1}^{m} P_i = 1. \qquad (4)$$

In case of the equations (1) - (3) for turbulent multiphase-phase flow, the function-indicator can be treated as a mathematical expectation, so that the solution of the equations (4) averaged by time on the time interval [0, T] are resulting in the functions-indicators of the phases. In case of two phases, the equations (4) with the corresponding initial conditions are as follows

$$\frac{dP_1}{dt} = -\lambda_{12} P_1 + \lambda_{21} P_2, \quad \frac{dP_2}{dt} = -\lambda_{21} P_2 + \lambda_{12} P_1, \quad P_1 + P_2 = 1; \quad t = 0, \quad P_1 = 1, \quad P_2 = 0. \qquad (5)$$

We can consider any other starting point for the phases, this is one of them taken in (5) just for example. The Cauchy problem (5) contains three equations for two functions, therefore it is over defined. Thus, we can solve a system of any two of the equations and then put the solution obtained into the third one.

The solution of the (5) is

$$P_1 = \frac{\lambda_{12} e^{-(\lambda_{12}+\lambda_{21})t} + \lambda_{21}}{\lambda_{12} + \lambda_{21}}. \qquad (6)$$

After averaging by the time it yields

$$P_1 = \frac{\lambda_{21}}{\lambda_{12} + \lambda_{21}} + \frac{\lambda_{12}}{\lambda_{12} + \lambda_{21}} \cdot \frac{1 - e^{-(\lambda_{12}+\lambda_{21})T}}{(\lambda_{12} + \lambda_{21})T}. \qquad (7)$$

With running of time it going to the following simpler expression

$$P_1 = \frac{\lambda_{21}}{\lambda_{12} + \lambda_{21}} = \frac{1}{\lambda_{12}/\lambda_{21} + 1} = \frac{1}{\gamma + 1}, \quad \gamma = \lambda_{12}/\lambda_{21}, \quad \frac{dP_1}{dz} = -\frac{1}{(\gamma+1)^2} \frac{d\gamma}{dz}$$

Obviously, function-indicator $B_1$ depends on the interval of averaging as far as the phases are interacting. By long time interval $P_1 = \lambda_{21}/(\lambda_{12} + \lambda_{21})$ so that it is determined by intensities of the phases' interaction. As far as at each point of the flow these coefficients are different, the function-indicator changes from point to point. For small time interval, it is got $B_1=1$ with accuracy to linear terms by $T$.

Each phase occupies some part of elementary volume of the heterogeneous medium: the volumetric contents of $N$ phases satisfy to the equation $\sum_{j=1}^{N} \alpha_j = 1$, the density of the medium is expressed through the real densities of phases by $\rho = \sum_{j=1}^{N} \rho_j \alpha_j$. At each point of the heterogeneous medium, $N$ parameters are defined relating to a continuum (densities, velocities, temperatures, etc.). The set of continuums, each of which corresponds to its phase and fills the same volume, is called the multi-speed continuum. Such approach is applied in the most multiphase methods, except just a few of them like the one in [12].

Using the above approach with the introduction of the probability of the phases at each point of a flow, all characteristics of the mixture $a^l(t)$ (mass, velocity, momentum, energy, etc.) of the corresponding characteristics of different phases $a_i^l(t)$ in a multiphase flow was proposed to be expressed as follows

$$a^l(t) = \sum_{i=1}^{m} P_i(t) a_i^l(t), \qquad (8)$$

where $P_i(t)$ was introduced as the above-stated and is connected with co-called function-indicator $B_i(t)$ by Nakorchevski [12]. Summation of the equations of the mix by all phases gives the equations of the heterogeneous medium taken as a uniform system, without an account of internal structure. Such model doesn't show features of interfacial interaction in the heterogeneous mix. Contrary to it, accounting an interaction of the phases' macroscopic inclusions results in the need for the account the conditions of joint deformation and movement of phases, influences of a form and the number of inclusions, their distributions in space, phase transformations, etc.

If physic-mechanical processes in the continua are rather precisely described by continuous or nearly everywhere continuous functions of coordinates and time, it is possible to replace the system of integral conservation equations to the corresponding differential equations. However, for the real continuous media prone to the external influences, the classical methods can be unacceptable owing to what the variation and numerical methods based on the use of the integral correlations are remaining useful in case of fractured fields and media if integration by Riemann is replaced by integration by Lebesgue. In a region of continuous or nearly everywhere continuous movement of a continuous medium, using Gauss-Ostrogradsky's formula, it is possible to pass from the integral balance equations to the differential equation array describing thermo-hydrodynamic processes in the heterogeneous medium taking into account the joint movement of phases and the interfacial mass, momentum, and energy exchange. The main obstacle in use of this system by mathematical modeling of heterogeneous media is caused by the need of a specification of laws of phases' interactions that is extremely difficult. The law of deformation of the heterogeneous medium depends not only on velocity fields, pressure, and temperature of phases; therefore, the determination of regularities of interfacial interaction even for special cases is a very complex challenge. And still, the accounting of fields' ruptures on the boundary interfaces is absolutely necessary for some practically important tasks.

At rather a weak manifestation of interfacial interaction in the heterogeneous media, for the description of the processes happening in it, the system of differential equations obtained from the conservation equations by each component of the heterogeneous mix through the summation on all mix may be used. But remarkably the balance equations of an impulse and energy depend on the relative movement of phases inside the heterogeneous mix. From (8) for two-phase flow is got

$$a^l(t) = P_1(t)a_1^l(t) + P_2(t)a_2^l(t) = P_1(t)a_1^l(t) + (1-P_1(t))a_2^l(t) = a_2^l(t) + P_1(t)(a_1^l(t) - a_2^l(t)). \quad (9)$$

## 3. The equation array of the mass, momentum and energy conservation in two-phase flow

### Non-stationary two-dimensional two-phase flow

According to the above-stated, the equation array for the two-phase flow of incompressible immiscible liquids without the twisting $(\partial \varphi / \partial r = 0)$ in the cylindrical coordinate system $(r, \varphi, z)$ is the following

$$\frac{\partial}{\partial r}\left\{\left[\rho_2 + P_1(\rho_1 - \rho_2)\right]\left[u_2 + P_1(u_1 - u_2)\right]\right\} + \frac{1}{r}\left[\rho_2 + P_1(\rho_1 - \rho_2)\right]\left[u_2 + P_1(u_1 - u_2)\right] +$$

$$+ \frac{\partial}{\partial z}\left\{\left[\rho_2 + P_1(\rho_1 - \rho_2)\right]\left[w_2 + P_1(w_1 - w_2)\right]\right\} + \frac{\partial}{\partial t}\left[\rho_2 + P_1(\rho_1 - \rho_2)\right] = 0,$$

$$\left\{\frac{\partial}{\partial t}\left[u_2 + P_1(u_1 - u_2)\right] + \frac{1}{2}\frac{\partial}{\partial r}\left[u_2 + P_1(u_1 - u_2)\right]^2 + \left[w_2 + P_1(w_1 - w_2)\right]\frac{\partial}{\partial z}\left[u_2 + P_1(u_1 - u_2)\right]\right\}\cdot$$

$$\cdot\left[\rho_2 + P_1(\rho_1 - \rho_2)\right] + \frac{\partial}{\partial r}\left[p_2 + P_1(p_1 - p_2)\right] = \frac{\partial}{\partial r}\left\{\left[\mu_2 + P_1(\mu_1 - \mu_2)\right]\frac{\partial}{\partial r}\left[u_2 + P_1(u_1 - u_2)\right]\right\} +$$

$$+ \frac{\partial}{\partial z}\left\{\left[\mu_2 + P_1(\mu_1 - \mu_2)\right]\frac{\partial}{\partial z}\left[u_2 + P_1(u_1 - u_2)\right]\right\} - \frac{1}{r^2}\left[\mu_2 + P_1(\mu_1 - \mu_2)\right]\left[u_2 + P_1(u_1 - u_2)\right] +$$

$$+ \frac{\partial}{r\partial r}\left[\mu_2 + P_1(\mu_1 - \mu_2)\right]\left[u_2 + P_1(u_1 - u_2)\right], \quad (10)$$

$$\left\{\frac{\partial}{\partial t}\left[w_2 + P_1(w_1 - w_2)\right] + \frac{1}{2}\frac{\partial}{\partial r}\left[w_2 + P_1(w_1 - w_2)\right]^2 + \left[u_2 + P_1(u_1 - u_2)\right]\frac{\partial}{\partial r}\left[w_2 + P_1(w_1 - w_2)\right]\right\}\cdot$$

$$\cdot\left[\rho_{2}+P_{1}(\rho_{1}-\rho_{2})\right]+\frac{\partial}{\partial z}\left[p_{2}+P_{1}(p_{1}-p_{2})\right]=\frac{\partial}{\partial r}\left\{\left[\mu_{2}+P_{1}(\mu_{1}-\mu_{2})\right]\frac{\partial}{\partial r}\left[w_{2}+P_{1}(w_{1}-w_{2})\right]\right\}+$$

$$+\frac{\partial}{\partial z}\left\{\left[\mu_{2}+P_{1}(\mu_{1}-\mu_{2})\right]\frac{\partial}{\partial z}\left[w_{2}+P_{1}(w_{1}-w_{2})\right]\right\}+\frac{\partial}{r\partial r}\left\{\left[\mu_{2}+P_{1}(\mu_{1}-\mu_{2})\right]\left[w_{2}+P_{1}(w_{1}-w_{2})\right]\right\}+$$

$$+\left[\rho_{2}+P_{1}(\rho_{1}-\rho_{2})\right]g,$$

$$\left[\rho_{2}+P_{1}(\rho_{1}-\rho_{2})\right]\left[c_{V2}+P_{1}(c_{V1}-c_{V2})\right]\left\{\frac{\partial}{\partial t}\left[T_{2}+P_{1}(T_{1}-T_{2})\right]+\left[u_{2}+P_{1}(u_{1}-u_{2})\right]\cdot\right.$$

$$\left.\cdot\frac{\partial}{\partial r}\left[T_{2}+P_{1}(T_{1}-T_{2})\right]+\left[w_{2}+P_{1}(w_{1}-w_{2})\right]\frac{\partial}{\partial z}\left[T_{2}+P_{1}(T_{1}-T_{2})\right]\right\}+\left[p_{2}+P_{1}(p_{1}-p_{2})\right]\cdot$$

$$\cdot\left\{\frac{\partial}{\partial r}\left[u_{2}+P_{1}(u_{1}-u_{2})\right]+\frac{1}{r}\left[u_{2}+P_{1}(u_{1}-u_{2})\right]+\frac{\partial}{\partial z}\left[w_{2}+P_{1}(w_{1}-w_{2})\right]\right\}=$$

$$=\frac{\partial}{\partial r}\left\{\left[\kappa_{2}+P_{1}(\kappa_{1}-\kappa_{2})\right]\frac{\partial}{\partial r}\left[T_{2}+P_{1}(T_{1}-T_{2})\right]\right\}+\frac{\partial}{r\partial r}\left\{\left[\kappa_{2}+P_{1}(\kappa_{1}-\kappa_{2})\right]\left[T_{2}+P_{1}(T_{1}-T_{2})\right]\right\}+$$

$$+\frac{\partial}{\partial z}\left\{\left[\kappa_{2}+P_{1}(\kappa_{1}-\kappa_{2})\right]\frac{\partial}{\partial z}\left[T_{2}+P_{1}(T_{1}-T_{2})\right]\right\}+\frac{1}{2}\left[\mu_{2}+P_{1}(\mu_{1}-\mu_{2})\right]\left\{\frac{\partial}{\partial z}\left[u_{2}+P_{1}(u_{1}-u_{2})\right]+\right.$$

$$\left.+\frac{\partial}{\partial r}\left[w_{2}+P_{1}(w_{1}-w_{2})\right]\right\}^{2},$$

where $r$ is the radial coordinate, $z$ is directed along the axis, $u$, $w$ are the corresponding velocity components, $p$ and $T$ are pressure and temperature; $c, \kappa, \mu$ are the heat capacity, heat conductivity, and dynamic viscosity coefficients, respectively. Here gravitational force is directed along the coordinate $z$, so that to act in the same direction as the momentum of a flow (liquid flow vertically down). If the liquid flow is going vertically up, the sign of the gravitational force must be negative, which we can account by the sign of the value $g$ (acceleration due to gravity).

### The equation array for stationary two-phase flow in cylindrical coordinate system

From the equation array (10) follows the next, for the case of constant densities of the phases and physical properties follows, accounting the expression (6) obtained for probabilities of the phases:

$$\left[u_{2}+P_{1}(u_{1}-u_{2})\right](\rho_{1}-\rho_{2})\frac{\partial P_{1}}{\partial r}+\left[\rho_{2}+P_{1}(\rho_{1}-\rho_{2})\right]\left\{\frac{\partial}{\partial r}\left[u_{2}+P_{1}(u_{1}-u_{2})\right]+\frac{u_{2}+P_{1}(u_{1}-u_{2})}{r}\right\}+$$

$$+\left[w_{2}+P_{1}(w_{1}-w_{2})\right](\rho_{1}-\rho_{2})\frac{\partial P_{1}}{\partial z}+\left[\rho_{2}+P_{1}(\rho_{1}-\rho_{2})\right]\frac{\partial}{\partial z}\left[w_{2}+P_{1}(w_{1}-w_{2})\right]+(\rho_{1}-\rho_{2})\frac{\partial P_{1}}{\partial t}=0,$$

$$\left\{\frac{\partial}{\partial t}\left[u_{2}+P_{1}(u_{1}-u_{2})\right]+\frac{1}{2}\frac{\partial}{\partial r}\left[u_{2}+P_{1}(u_{1}-u_{2})\right]^{2}+\left[w_{2}+P_{1}(w_{1}-w_{2})\right]\frac{\partial}{\partial z}\left[u_{2}+P_{1}(u_{1}-u_{2})\right]\right\}\cdot$$

$$\cdot\left[\rho_{2}+P_{1}(\rho_{1}-\rho_{2})\right]+\frac{\partial}{\partial r}\left[p_{2}+P_{1}(p_{1}-p_{2})\right]=(\mu_{1}-\mu_{2})\frac{\partial P_{1}}{\partial r}\frac{\partial}{\partial r}\left[u_{2}+P_{1}(u_{1}-u_{2})\right]+\left[\mu_{2}+P_{1}(\mu_{1}-\mu_{2})\right]\cdot$$

$$\cdot\frac{\partial^{2}}{\partial r^{2}}\left[u_{2}+P_{1}(u_{1}-u_{2})\right]+(\mu_{1}-\mu_{2})\frac{\partial P_{1}}{\partial z}\frac{\partial}{\partial z}\left[u_{2}+P_{1}(u_{1}-u_{2})\right]-\frac{\mu_{2}+P_{1}(\mu_{1}-\mu_{2})}{r^{2}}\left[u_{2}+P_{1}(u_{1}-u_{2})\right]+$$

$$+\frac{(\mu_1-\mu_2)}{r}\left[u_2+P_1(u_1-u_2)\right]\frac{\partial P_1}{\partial r}+\frac{(\mu_1-\mu_2)}{r}\left[u_2+P_1(u_1-u_2)\right]\frac{\partial P_1}{\partial r}+$$

$$+\left[\mu_2+P_1(\mu_1-\mu_2)\right]\frac{\partial^2}{\partial z^2}\left[u_2+P_1(u_1-u_2)\right]+\frac{1}{r}\left[\mu_2+P_1(\mu_1-\mu_2)\right]\frac{\partial}{\partial r}\left[u_2+P_1(u_1-u_2)\right],$$

$$\left\{\frac{\partial}{\partial t}\left[w_2+P_1(w_1-w_2)\right]+\frac{1}{2}\frac{\partial}{\partial r}\left[w_2+P_1(w_1-w_2)\right]^2+\left[u_2+P_1(u_1-u_2)\right]\frac{\partial}{\partial r}\left[w_2+P_1(w_1-w_2)\right]\right\}\cdot$$

$$\cdot\left[\rho_2+P_1(\rho_1-\rho_2)\right]+\frac{\partial}{\partial z}\left[p_2+P_1(p_1-p_2)\right]=(\mu_1-\mu_2)\frac{\partial P_1}{\partial r}\frac{\partial}{\partial r}\left[w_2+P_1(w_1-w_2)\right]+$$

$$+\left[\mu_2+P_1(\mu_1-\mu_2)\right]\frac{\partial^2}{\partial r^2}\left[w_2+P_1(w_1-w_2)\right]+(\mu_1-\mu_2)\frac{\partial P_1}{\partial z}\frac{\partial}{\partial z}\left[w_2+P_1(w_1-w_2)\right]+$$

$$+\left[\mu_2+P_1(\mu_1-\mu_2)\right]\frac{\partial^2}{\partial z^2}\left[w_2+P_1(w_1-w_2)\right]+\frac{\partial}{r\partial r}\left\{\left[\mu_2+P_1(\mu_1-\mu_2)\right]\left[w_2+P_1(w_1-w_2)\right]\right\}+$$

$$+\left[\rho_2+P_1(\rho_1-\rho_2)\right]g, \qquad (11)$$

$$\left[\rho_2+P_1(\rho_1-\rho_2)\right]\left[c_{V2}+P_1(c_{V1}-c_{V2})\right]\left\{\frac{\partial}{\partial t}\left[T_2+P_1(T_1-T_2)\right]+\left[u_2+P_1(u_1-u_2)\right]\cdot\right.$$

$$\left.\cdot\frac{\partial}{\partial r}\left[T_2+P_1(T_1-T_2)\right]+\left[w_2+P_1(w_1-w_2)\right]\frac{\partial}{\partial z}\left[T_2+P_1(T_1-T_2)\right]\right\}+\left[p_2+P_1(p_1-p_2)\right]\cdot$$

$$\cdot\left\{\frac{\partial}{\partial r}\left[u_2+P_1(u_1-u_2)\right]+\frac{1}{r}\left[u_2+P_1(u_1-u_2)\right]+\frac{\partial}{\partial z}\left[w_2+P_1(w_1-w_2)\right]\right\}=$$

$$=(\kappa_1-\kappa_2)\frac{\partial P_1}{\partial r}\frac{\partial}{\partial r}\left[T_2+P_1(T_1-T_2)\right]+\left[\kappa_2+P_1(\kappa_1-\kappa_2)\right]\frac{\partial^2}{\partial r^2}\left[T_2+P_1(T_1-T_2)\right]+$$

$$+\frac{\kappa_1-\kappa_2}{r}\left[T_2+P_1(T_1-T_2)\right]\frac{\partial P_1}{\partial r}+\frac{\kappa_2+P_1(\kappa_1-\kappa_2)}{r}\frac{\partial}{\partial r}\left[T_2+P_1(T_1-T_2)\right]+$$

$$+(\kappa_1-\kappa_2)\frac{\partial}{\partial z}\left[T_2+P_1(T_1-T_2)\right]\frac{\partial P_1}{\partial z}+\left[\kappa_2+P_1(\kappa_1-\kappa_2)\right]\frac{\partial^2}{\partial z^2}\left[T_2+P_1(T_1-T_2)\right]+\frac{1}{2}\left[\mu_2+P_1(\mu_1-\mu_2)\right]\cdot$$

$$\cdot\left\{\frac{\partial}{\partial z}\left[u_2+P_1(u_1-u_2)\right]+\frac{\partial}{\partial r}\left[w_2+P_1(w_1-w_2)\right]\right\}^2.$$

**The one-dimensional stationary two-phase flow in cylindrical coordinate system**

The simplest one-dimensional flow is described according to the equations (11) with the following system of equations ($\partial/\partial r=0$, the channel is narrow and characteristics are mainly depending on the coordinate $z$ along the axis of the channel):

$$\frac{u_2+P_1(u_1-u_2)}{r}+\frac{w_2+P_1(w_1-w_2)}{\rho_2+P_1(\rho_1-\rho_2)}(\rho_1-\rho_2)\frac{\partial P_1}{\partial z}+\frac{\partial}{\partial z}\left[w_2+P_1(w_1-w_2)\right]+\frac{(\rho_1-\rho_2)}{\rho_2+P_1(\rho_1-\rho_2)}\frac{\partial P_1}{\partial t}=0,$$

$$\frac{\partial^2}{\partial z^2}\left[u_2+P_1(u_1-u_2)\right]+\left\{\frac{\mu_1-\mu_2}{\mu_2+P_1(\mu_1-\mu_2)}\frac{\partial P_1}{\partial z}-\frac{\rho_2+P_1(\rho_1-\rho_2)}{\mu_2+P_1(\mu_1-\mu_2)}\left[w_2+P_1(w_1-w_2)\right]\right\}\cdot$$

$$\cdot \frac{\partial}{\partial z}\left[u_2 + P_1(u_1 - u_2)\right] = \frac{u_2 + P_1(u_1 - u_2)}{r^2} + \frac{\rho_2 + P_1(\rho_1 - \rho_2)}{\mu_2 + P_1(\mu_1 - \mu_2)} \frac{\partial}{\partial t}\left[u_2 + P_1(u_1 - u_2)\right], \quad (12)$$

$$\frac{\partial^2}{\partial z^2}\left[w_2 + P_1(w_1 - w_2)\right] + \left\{\frac{\mu_1 - \mu_2}{\mu_2 + P_1(\mu_1 - \mu_2)} \frac{\partial P_1}{\partial z}\right\} \frac{\partial}{\partial z}\left[w_2 + P_1(w_1 - w_2)\right] =$$

$$= \frac{\rho_2 + P_1(\rho_1 - \rho_2)}{\mu_2 + P_1(\mu_1 - \mu_2)} \frac{\partial}{\partial t}\left[w_2 + P_1(w_1 - w_2)\right] + \frac{1}{\mu_2 + P_1(\mu_1 - \mu_2)} \frac{\partial}{\partial z}\left[p_2 + P_1(p_1 - p_2)\right] - \frac{\rho_2 + P_1(\rho_1 - \rho_2)}{\mu_2 + P_1(\mu_1 - \mu_2)} g,$$

$$\left[c_{V2} + P_1(c_{V1} - c_{V2})\right]\left\{\frac{\partial}{\partial t}\left[T_2 + P_1(T_1 - T_2)\right] + \left[w_2 + P_1(w_1 - w_2)\right]\frac{\partial}{\partial z}\left[T_2 + P_1(T_1 - T_2)\right]\right\} +$$

$$+ \frac{p_2 + P_1(p_1 - p_2)}{\rho_2 + P_1(\rho_1 - \rho_2)}\left\{\frac{1}{r}\left[u_2 + P_1(u_1 - u_2)\right] + \frac{\partial}{\partial z}\left[w_2 + P_1(w_1 - w_2)\right]\right\} = \frac{\kappa_2 + P_1(\kappa_1 - \kappa_2)}{\rho_2 + P_1(\rho_1 - \rho_2)} \frac{\partial^2}{\partial z^2}\left[T_2 + P_1(T_1 - T_2)\right] +$$

$$+ \frac{\kappa_1 - \kappa_2}{\rho_2 + P_1(\rho_1 - \rho_2)} \frac{\partial}{\partial z}\left[T_2 + P_1(T_1 - T_2)\right]\frac{\partial P_1}{\partial z} + \frac{\mu_2 + P_1(\mu_1 - \mu_2)}{2\left[\rho_2 + P_1(\rho_1 - \rho_2)\right]}\left\{\frac{\partial}{\partial z}\left[u_2 + P_1(u_1 - u_2)\right]\right\}^2.$$

The stationary regime of the above flow described by the equation array (12) is the following

$$\frac{d^2}{dz^2}\left[u_2 + P_1(u_1 - u_2)\right] + \left\{\frac{\mu_1 - \mu_2}{\mu_2 + P_1(\mu_1 - \mu_2)} \frac{dP_1}{dz} - \frac{\rho_2 + P_1(\rho_1 - \rho_2)}{\mu_2 + P_1(\mu_1 - \mu_2)}\left[w_2 + P_1(w_1 - w_2)\right]\right\}\frac{d}{dz}\left[u_2 + P_1(u_1 - u_2)\right] =$$

$$= \frac{u_2 + P_1(u_1 - u_2)}{r^2}, \quad \frac{u_2 + P_1(u_1 - u_2)}{r} + \frac{w_2 + P_1(w_1 - w_2)}{\rho_2 + P_1(\rho_1 - \rho_2)}(\rho_1 - \rho_2)\frac{dP_1}{dz} + \frac{d}{dz}\left[w_2 + P_1(w_1 - w_2)\right] = 0,$$

$$\frac{d^2}{dz^2}\left[w_2 + P_1(w_1 - w_2)\right] + \left\{\frac{\mu_1 - \mu_2}{\mu_2 + P_1(\mu_1 - \mu_2)} \frac{dP_1}{dz}\right\}\frac{d}{dz}\left[w_2 + P_1(w_1 - w_2)\right] = \frac{1}{\mu_2 + P_1(\mu_1 - \mu_2)}$$

$$\cdot \frac{d}{dz}\left[p_2 + P_1(p_1 - p_2)\right] - \frac{\rho_2 + P_1(\rho_1 - \rho_2)}{\mu_2 + P_1(\mu_1 - \mu_2)} g, \quad (13)$$

$$\left[c_{V2} + P_1(c_{V1} - c_{V2})\right]\left[w_2 + P_1(w_1 - w_2)\right]\frac{d}{dz}\left[T_2 + P_1(T_1 - T_2)\right] + \frac{p_2 + P_1(p_1 - p_2)}{\rho_2 + P_1(\rho_1 - \rho_2)}\cdot$$

$$\cdot\left\{\frac{1}{r}\left[u_2 + P_1(u_1 - u_2)\right] + \frac{d}{dz}\left[w_2 + P_1(w_1 - w_2)\right]\right\} = \frac{\kappa_1 - \kappa_2}{\rho_2 + P_1(\rho_1 - \rho_2)} \frac{d}{dz}\left[T_2 + P_1(T_1 - T_2)\right]\frac{dP_1}{dz} +$$

$$+ \frac{\kappa_2 + P_1(\kappa_1 - \kappa_2)}{\rho_2 + P_1(\rho_1 - \rho_2)} \frac{d^2}{dz^2}\left[T_2 + P_1(T_1 - T_2)\right] + \frac{\mu_2 + P_1(\mu_1 - \mu_2)}{2\left[\rho_2 + P_1(\rho_1 - \rho_2)\right]}\left\{\frac{d}{dz}\left[u_2 + P_1(u_1 - u_2)\right]\right\}^2.$$

Also, it can be presented in the next form

$$\frac{d^2}{dz^2}\left[u_2 + P_1(u_1 - u_2)\right] + \left\{\frac{d}{dz}\ln\left[\mu_2 + P_1(\mu_1 - \mu_2)\right] - \frac{\rho_2 + P_1(\rho_1 - \rho_2)}{\mu_2 + P_1(\mu_1 - \mu_2)}\left[w_2 + P_1(w_1 - w_2)\right]\right\}\cdot$$

$$\cdot \frac{d}{dz}\left[u_2 + P_1(u_1 - u_2)\right] = \frac{u_2 + P_1(u_1 - u_2)}{r^2}, \quad \frac{u_2 + P_1(u_1 - u_2)}{r} + \frac{w_2 + P_1(w_1 - w_2)}{\rho_2 + P_1(\rho_1 - \rho_2)}(\rho_1 - \rho_2)\frac{dP_1}{dz} +$$

$$+ \frac{d}{dz}\left[w_2 + P_1(w_1 - w_2)\right] = 0, \quad \frac{d^2}{dz^2}\left[w_2 + P_1(w_1 - w_2)\right] + \left\{\frac{\mu_1 - \mu_2}{\mu_2 + P_1(\mu_1 - \mu_2)}\frac{dP_1}{dz}\right\}\frac{d}{dz}\left[w_2 + P_1(w_1 - w_2)\right] =$$

$$= \frac{1}{\mu_2 + P_1(\mu_1 - \mu_2)}\frac{d}{dz}\left[p_2 + P_1(p_1 - p_2)\right] - \frac{\rho_2 + P_1(\rho_1 - \rho_2)}{\mu_2 + P_1(\mu_1 - \mu_2)}g, \left[c_{V2} + P_1(c_{V1} - c_{V2})\right]\left[w_2 + P_1(w_1 - w_2)\right] \cdot$$

$$\cdot \frac{d}{dz}\left[T_2 + P_1(T_1 - T_2)\right] + \frac{p_2 + P_1(p_1 - p_2)}{\rho_2 + P_1(\rho_1 - \rho_2)}\left\{\frac{1}{r}\left[u_2 + P_1(u_1 - u_2)\right] + \frac{d}{dz}\left[w_2 + P_1(w_1 - w_2)\right]\right\} =$$

$$= \frac{\kappa_1 - \kappa_2}{\rho_2 + P_1(\rho_1 - \rho_2)}\frac{d}{dz}\left[T_2 + P_1(T_1 - T_2)\right]\frac{dP_1}{dz} + \quad (14)$$

$$+ \frac{\kappa_2 + P_1(\kappa_1 - \kappa_2)}{\rho_2 + P_1(\rho_1 - \rho_2)}\frac{d^2}{dz^2}\left[T_2 + P_1(T_1 - T_2)\right] + \frac{\mu_2 + P_1(\mu_1 - \mu_2)}{2\left[\rho_2 + P_1(\rho_1 - \rho_2)\right]}\left\{\frac{d}{dz}\left[u_2 + P_1(u_1 - u_2)\right]\right\}^2.$$

**Dimensionless form of the one-dimensional stationary two-phase flow**

The equation array is presented in a dimensionless form for generalization of the solution and its analysis. The scales for the velocity, length, pressure and temperature are taken as follows: $u_0$, $R_0$, $p_0$, $T_0$ and for the dimensionless parameters we keep the same assignments as before for the dimensional ones. The dimensionless system is presented in the form:

$$\left\{\frac{d}{dz}\ln\left[\mu_{21} + P_1(1 - \mu_{21})\right] - \frac{\rho_{21} + P_1(1 - \rho_{21})}{\mu_{21} + P_1(1 - \mu_{21})}\left[w_2 + P_1(w_1 - w_2)\right]\text{Re}\right\}\frac{d}{dz}\left[u_2 + P_1(u_1 - u_2)\right] +$$

$$+ \frac{d^2}{dz^2}\left[u_2 + P_1(u_1 - u_2)\right] = \frac{u_2 + P_1(u_1 - u_2)}{r^2}, \quad \frac{w_2 + P_1(w_1 - w_2)}{\rho_{21} + P_1(1 - \rho_{21})}(1 - \rho_{21})\frac{dP_1}{dz} + \frac{d}{dz}\left[w_2 + P_1(w_1 - w_2)\right] +$$

$$+ \frac{u_2 + P_1(u_1 - u_2)}{r} = 0, \quad \frac{d^2}{dz^2}\left[w_2 + P_1(w_1 - w_2)\right] + \frac{d\ln\left[\mu_{21} + P_1(1 - \mu_{21})\right]}{dz}\frac{d}{dz}\left[w_2 + P_1(w_1 - w_2)\right] =$$

$$= \frac{Eu \cdot \text{Re}}{\mu_{21} + P_1(1 - \mu_{21})}\frac{d}{dz}\left[p_2 + P_1(p_1 - p_2)\right] - \frac{\rho_{21} + P_1(1 - \rho_{21})}{\mu_{21} + P_1(1 - \mu_{21})}\frac{Ga}{\text{Re}}, \quad (15)$$

$$\left[c_{V21} + P_1(1 - c_{V21})\right]\left[w_2 + P_1(w_1 - w_2)\right]\frac{d}{dz}\left[T_2 + P_1(T_1 - T_2)\right] + c_{pv1}\frac{p_2 + P_1(p_1 - p_2)}{\rho_{21} + P_1(1 - \rho_{21})}\cdot Eu \cdot Ec \cdot$$

$$\cdot\left\{\frac{1}{r}\left[u_2 + P_1(u_1 - u_2)\right] + \frac{d}{dz}\left[w_2 + P_1(w_1 - w_2)\right]\right\} = \frac{(1 - \kappa_{21})c_{pv1}/Pe}{\rho_{21} + P_1(1 - \rho_{21})}\frac{d}{dz}\left[T_2 + P_1(T_1 - T_2)\right]\frac{dP_1}{dz} +$$

$$+ \frac{\kappa_{21} + P_1(1 - \kappa_{21})}{\rho_{21} + P_1(1 - \rho_{21})}\frac{d^2}{dz^2}\left[T_2 + P_1(T_1 - T_2)\right]\frac{c_{pv1}}{Pe} + \frac{\left[\mu_{21} + P_1(1 - \mu_{21})\right]c_{pv1}}{2\left[\rho_{21} + P_1(1 - \rho_{21})\right]}\left\{\frac{d}{dz}\left[u_2 + P_1(u_1 - u_2)\right]\right\}^2 \cdot \frac{Ec}{\text{Re}}.$$

Here are: $Ec = u_0^2/(c_{p1}T_0)$, $Ga = gR_0^3/v_1^2$, $Eu = p_0/(\rho_1 u_0^2)$, $\text{Re} = u_0 R_0/v_1$, $Pe = u_0 R_0/a_1$, $v_1 = \mu_1/\rho_1$, $\kappa_{21} = \kappa_2/\kappa_1$, $\rho_{21} = \rho_2/\rho_1$, $\mu_{21} = \mu_2/\mu_1$, $a_1 = \mu_1/(\rho_1 c_{p1})$, $c_{pv1} = c_{p1}/c_{v1}$. The Eckert number $Ec$ plays an important role, representing the ratio of kinetic energy at the wall to the specific enthalpy of a fluid, the Galileo number is a dimensionless group representing a ratio of the forces present in the flow of viscous fluids (gravitational*momentum force/ viscous force), $Eu$ – the Euler criterion (characteristic pressure $p_0$ to kinetic energy of a flow $\rho_1 u_0^2$), $Re$ – the Reynolds number, $Pe$ – the Peclet number.

## 4. Analysis of the mathematical model derived for the case of 1-D stationary two-phase flow

### Parameters of the two-phase mixture and distribution of probabilities of the phases

System (15) contains 4 differential equations for 9 functions: $P_1$, $u_1$, $u_2$, $w_1$, $w_2$, $T_1$, $T_2$, $p_1$, $p_2$. Thus, the system is not closed because the multiphase system is having no closing relations, which are the mass, momentum and energy interaction between the phases. The last problem is the main obstacle in any multiphase system. It is not solved in general statement, only for specific cases through the structural approach. The system (15) is written for the heterogeneous mixture, which does not "feel" interaction of the phases. It can be used for the study of the multiphase system having some experimental data allowing the closing (15) for further its solution. We can propose the following methodology for solving this task. First let us consider the known temperature distribution for the mixture $\bar{T} = T_2 + P_1(T_1 - T_2) = \alpha_0 + \alpha_1 z + \alpha_2 z^2$, and let us consider system (15) searching solution for 4 functions: $\bar{w} = w_2 + P_1(w_1 - w_2)$, $\bar{u} = u_2 + P_1(u_1 - u_2)$, $\bar{p} = p_2 + P_1(p_1 - p_2)$, $P_1$. The function $P_1(z)$ is only one showing the multiphase nature of the system. There are also a few terms depending on $r$. As far as we assumed that solution depends only on $z$, we consider that only $P_1(r, z)$ is causing this. Thus, from (15) follows:

$$\left(\frac{d}{dz}\ln\bar{\mu} - \frac{\bar{\rho}}{\bar{\mu}}\bar{w}\operatorname{Re}\right)\frac{d\bar{u}}{dz} + \frac{d^2\bar{u}}{dz^2} = \frac{\bar{u}}{r^2}, \quad \frac{\bar{w}}{\bar{\rho}}(1-\rho_{21})\frac{dP_1}{dz} + \frac{d\bar{w}}{dz} + \frac{\bar{u}}{r} = 0, \quad \frac{d^2\bar{w}}{dz^2} + \frac{d\ln\bar{\mu}}{dz}\frac{d\bar{w}}{dz} = \frac{Eu\cdot\operatorname{Re}}{\bar{\mu}}\frac{d\bar{p}}{dz} - \frac{\bar{\rho}}{\bar{\mu}}\frac{Ga}{\operatorname{Re}},$$

$$\bar{c}_{21}\bar{w}(\alpha_1 + 2\alpha_2 z) + c_{pv1}\frac{\bar{p}}{\bar{\rho}}Eu\cdot Ec\left(\frac{\bar{u}}{r} + \frac{d\bar{w}}{dz}\right) = \frac{1-\kappa_{21}}{\bar{\rho}Pe}c_{pv1}(\alpha_1 + 2\alpha_2 z)\frac{dP_1}{dz} + 2\alpha_2\frac{\bar{\kappa}}{\bar{\rho}}\frac{c_{pv1}}{Pe} + \frac{\bar{\mu}c_{pv1}}{2\bar{\rho}}\left(\frac{d\bar{u}}{dz}\right)^2\frac{Ec}{\operatorname{Re}},$$

$$\bar{\mu} = \mu_{21} + P_1(1 - \mu_{21}), \quad \bar{c}_{21} = c_{V21} + P_1(1 - c_{V21}), \quad \bar{\kappa} = \kappa_{21} + P_1(1 - \kappa_{21}), \quad \bar{\rho} = \rho_{21} + P_1(1 - \rho_{21}). \tag{16}$$

First, let us consider strictly one-dimensional system, so that $\bar{u} = 0$, and account temperature effect in a simplest way as for the linear function by $z$ ($\alpha_2 = 0$), then the equation array (16) yields:

$$\frac{(\rho_{21} - 1)\bar{w}}{\rho_{21} + P_1(1-\rho_{21})}\frac{dP_1}{dz} = \frac{d\bar{w}}{dz}, \quad \frac{d^2\bar{w}}{dz^2} + \frac{d\ln\bar{\mu}}{dz}\frac{d\bar{w}}{dz} = \frac{Eu\cdot\operatorname{Re}}{\bar{\mu}}\frac{d\bar{p}}{dz} - \frac{\bar{\rho}}{\bar{\mu}}\frac{Ga}{\operatorname{Re}},$$

$$\bar{c}_{21}\bar{w}\alpha_1 + c_{pv1}\frac{\bar{p}}{\bar{\rho}}Eu\cdot Ec\frac{d\bar{w}}{dz} = \frac{1-\kappa_{21}}{\bar{\rho}Pe}c_{pv1}\alpha_1\frac{dP_1}{dz}, \tag{17}$$

Now from the first equation of the system (17) follows the next

$$\frac{1}{\bar{w}}\frac{d\bar{w}}{dz} = \frac{(\rho_{21} - 1)}{\rho_{21} + P_1(1-\rho_{21})}\frac{dP_1}{dz}, \quad \rightarrow \frac{d\bar{w}}{\bar{w}} = \frac{(\rho_{21} - 1)dP_1}{\rho_{21} + P_1(1-\rho_{21})}, \quad \rightarrow \ln\bar{w} + \ln[\rho_{21} + P_1(1-\rho_{21})] = const,$$

where from considering the following boundary condition:

$$z=0, \quad \bar{w}_0 = w_{20} + P_{10}(w_{10} - w_{20}), \quad \bar{\rho}_0 = \rho_{21} + P_{10}(1-\rho_{21}). \tag{18}$$

Now the final expression from the above equations with account of the boundary condition (18) is

$$\bar{w} = \frac{\bar{w}_0 \bar{\rho}_0}{\rho_{21} + P_1(1-\rho_{21})}, \quad \text{or} \quad P_1 = \frac{1}{1-\rho_{21}}\left(\frac{\bar{w}_0 \bar{\rho}_0}{\bar{w}} - \rho_{21}\right), \quad \frac{dP_1}{dz} = -\frac{\bar{w}_0 \bar{\rho}_0}{(1-\rho_{21})\bar{w}^2}\frac{d\bar{w}}{dz}. \tag{19}$$

Zero index means that values are taken by $z=0$. From the (19) we can get explicit expression for the probability of the first phase $P_1$ with account of the $\bar{w} = w_2 + P_1(w_1 - w_2)$:

$$(1-\rho_{21})(w_1 - w_2)P_1^2 + \left[(1-\rho_{21})w_2 + \rho_{21}(w_1 - w_2)\right]P_1 + \rho_{21}w_2 - \bar{\rho}_0\bar{w}_0 = 0. \quad (20)$$

### The correctness analysis of the obtained equation array and probability function

Analysis of the equation (20) shows that for the one-speed two-phase flow when $w_1 = w_2 = w$, phase probability is expressed $P_1 = P_{10}w_0/w$, where from the condition $P_1 < 1$ is satisfied by $w > P_{10}w_0$, so that if the first phase is presented in large amount in a two-phase flow, the flow velocity cannot fall in such flow, it must grow with a coordinate increasing involving a second phase. But if a first phase is far from maximal value (e.g. close to 0), then with decreasing of a flow velocity the amount of the first phase in a heterogeneous mixture is growing. Then for the two-phase flow of nearly the same densities ($\rho_{21} = 1$) we get from the condition $0 < P_1 < 1$ the following requirements: $w_1 < \bar{\rho}_0\bar{w}_0$, $w_2 > \bar{\rho}_0\bar{w}_0$, or the other case $w_1 > \bar{\rho}_0\bar{w}_0$, $w_2 < \bar{\rho}_0\bar{w}_0$.

A general solution of the equation (20) is as follows

$$(P_1)_{1,2} = \frac{(\rho_{21} - 1)w_2 + \rho_{21}(w_2 - w_1) \pm \sqrt{(\rho_{21}w_1 - w_2)^2 + 4\bar{\rho}_0\bar{w}_0(1-\rho_{21})(w_1 - w_2)}}{2(1-\rho_{21})(w_1 - w_2)}, \quad (21)$$

where from $D = (\rho_{21}w_1 - w_2)^2 - 4\bar{\rho}_0\bar{w}_0(1-\rho_{21})(w_2 - w_1)$ is first of all analyzed in detail because we need only the real values of $P_1$ ($D \geq 0$): $(\rho_{21}w_1 - w_2)^2 \geq 4\bar{\rho}_0\bar{w}_0(1-\rho_{21})(w_2 - w_1)$. Obviously, it is always satisfied by $\rho_{21} \leq 1$, $w_2 \leq w_1$ (lighter phase has lower velocity), and $\rho_{21} \geq 1$, $w_2 \geq w_1$ (heavier phase has higher velocity). But in case of the two-phase jet when lighter liquid is going from the nozzle into surroundings with heavier liquid in a rest being ejected by the first phase [12-14] it is not satisfied. Therefore $(\rho_{21}w_1 - w_2)^2 \geq 4\bar{\rho}_0\bar{w}_0(1-\rho_{21})(w_2 - w_1)$ for the 2 cases: $\rho_{21} \leq 1$, $w_2 \geq w_1$ and $\rho_{21} \geq 1$, $w_2 \leq w_1$. We solve the last inequality with regard to velocity components having the other parameters stated:

$$(\rho_{21}w_1 - w_2)^2 \geq 4\bar{\rho}_0\bar{w}_0(1-\rho_{21})(w_2 - w_1), \quad w_2 \leq (w_2)_1, \quad w_2 \geq (w_2)_2;$$
$$(w_2)_{1,2} = \rho_{21}w_1 + 2\bar{\rho}_0\bar{w}_0(1-\rho_{21}) \pm 2|1-\rho_{21}|\sqrt{\bar{\rho}_0\bar{w}_0(\bar{\rho}_0\bar{w}_0 - w_1)}. \quad (22)$$

Here the square root must be real, therefore $w_1 \leq \bar{\rho}_0\bar{w}_0$. And from (22) follows

$$\rho_{21} < 1, \quad w_2 \leq \rho_{21}w_1 + 2(1-\rho_{21})\bar{\rho}_0\bar{w}_0\left(1 - \sqrt{1 - \frac{w_1}{\bar{\rho}_0\bar{w}_0}}\right), \quad w_2 \geq \rho_{21}w_1 + 2(1-\rho_{21})\bar{\rho}_0\bar{w}_0\left(1 + \sqrt{1 - \frac{w_1}{\bar{\rho}_0\bar{w}_0}}\right);$$

$$\rho_{21} > 1, \quad w_2 \leq \rho_{21}w_1 + 2(1-\rho_{21})\bar{\rho}_0\bar{w}_0\left(1 + \sqrt{1 - \frac{w_1}{\bar{\rho}_0\bar{w}_0}}\right), \quad w_2 \geq \rho_{21}w_1 + 2(1-\rho_{21})\bar{\rho}_0\bar{w}_0\left(1 - \sqrt{1 - \frac{w_1}{\bar{\rho}_0\bar{w}_0}}\right). \quad (23)$$

Considering the two-phase flow in a positive direction by $z$, we have to conclude that the conditions (23) must be satisfied with the positive values of $w_2$, $w_1$.

### Statement the Cauchy problem for the numerical solution of the 1-D equation array

Then from the other two equations of the system (17) we can get

$$\frac{d^2\bar{w}}{dz^2} = \frac{Eu \cdot \mathrm{Re}(1-\rho_{21})\bar{w}}{(\mu_{21} - \rho_{21})\bar{w} + (1-\mu_{21})\bar{w}_0\bar{\rho}_0}\frac{d\bar{p}}{dz} - \frac{Ga}{\mathrm{Re}}\frac{(1-\rho_{21})\bar{w}_0\bar{\rho}_0}{(\mu_{21} - \rho_{21})\bar{w} + (1-\mu_{21})\bar{w}_0\bar{\rho}_0} +$$

$$-\frac{(1-\rho_{21})\overline{w}_0\overline{\rho}_0}{\left[(\mu_{21}-\rho_{21})\overline{w}+(1-\mu_{21})\overline{w}_0\overline{\rho}_0\right]\overline{w}}\left(\frac{d\overline{w}}{dz}\right)^2, \quad P_1 = \frac{1}{1-\rho_{21}}\left(\frac{\overline{w}_0\overline{\rho}_0}{\overline{w}}-\rho_{21}\right) \tag{24}$$

$$\overline{p}\frac{d\overline{w}}{dz} = \frac{\alpha_1(1-\kappa_{21})\overline{w}_0\overline{\rho}_0}{Pe\cdot Eu\cdot Ec(1-\rho_{21})}\left(-\frac{1}{\overline{w}^2}\frac{d\overline{w}}{dz}\right) - \frac{\alpha_1\overline{w}_0\overline{\rho}_0}{Eu\cdot Ec\cdot c_{pv1}(1-\rho_{21})\overline{w}}\left[(c_{V21}-\rho_{21})\overline{w}+(1-c_{V21})\overline{w}_0\overline{\rho}_0\right].$$

The equations above are better presented in the following standard form

$$\frac{d\overline{p}}{dz} = \frac{(\mu_{21}-\rho_{21})\overline{w}+(1-\mu_{21})\overline{w}_0\overline{\rho}_0}{Eu\cdot Re(1-\rho_{21})\overline{w}}\frac{d^2\overline{w}}{dz^2} + \frac{\overline{w}_0\overline{\rho}_0}{Eu\cdot Re\,\overline{w}}\left[\frac{Ga}{Re}+\frac{1}{\overline{w}}\left(\frac{d\overline{w}}{dz}\right)^2\right], \tag{25}$$

$$\frac{d\overline{w}}{dz} = Pe\cdot\frac{\alpha_1\overline{w}_0\overline{\rho}_0\overline{w}\left[(c_{V21}-\rho_{21})\overline{w}+(1-c_{V21})\overline{w}_0\overline{\rho}_0\right]}{c_{pv1}\left[\overline{p}\overline{w}^2Pe\cdot Eu\cdot Ec(1-\rho_{21})+\alpha_1(1-\kappa_{21})\overline{w}_0\overline{\rho}_0\right]}, \quad P_1 = \frac{1}{1-\rho_{21}}\left(\frac{\overline{w}_0\overline{\rho}_0}{\overline{w}}-\rho_{21}\right).$$

$$z=0, \quad \overline{w}_0 = w_{20} + P_{10}(w_{10}-w_{20}), \quad \overline{\rho}_0 = \rho_{21}+P_{10}(1-\rho_{21}), \quad \overline{p}=1. \tag{26}$$

The Cauchy problem (25), (26) can be solved numerically by the dimensionless criteria $Eu$, $Re$, $Ga$, $Pe$, $Ec$ specified. It will give the pressure and velocity of the two-phase mixture and probability of the first phase $P_1$. Afterward, by these parameters, from the expressions $\overline{w} = w_2 + P_1(w_1-w_2)$, $\overline{p} = p_2 + P_1(p_1-p_2)$, $\overline{T} = T_2 + P_1(T_1-T_2)$ we can estimate the influence of interacting phases having the function $P_1$.

**Comparison of the phase probability from the fluid flow and from the Kolmogorov theorem**

Reminded the obtained in (7) stationary function $P_1$ for the considered at the beginning of this paper Kolmogorov theorem and compared it with the solution above yields:

$$P_1 = \frac{\lambda_{21}}{\lambda_{12}+\lambda_{21}}, \quad P_1 = \frac{1}{1-\rho_{21}}\left(\frac{\overline{w}_0\overline{\rho}_0}{\overline{w}}-\rho_{21}\right), \quad \text{or} \quad P_1 = \frac{1}{1+\gamma}, \quad \gamma = \frac{\lambda_{12}}{\lambda_{21}}.$$

Then consider formally the correspondence of them. Let us determine from comparison of these expressions the values of $\lambda_{12}, \lambda_{21}$, which reflect interaction of the phases ($\lambda_{12}$ - intensity of transition from phase 1 to phase 2 and, inversely, $\lambda_{21}$ - from phase 2 to phase 1). We get the following

$$\lambda_{21} = \overline{w}_0\overline{\rho}_0 - \rho_{21}\overline{w}, \quad \lambda_{12} = \overline{w} - \overline{w}_0\overline{\rho}_0, \quad \gamma = \frac{\lambda_{12}}{\lambda_{21}} = \frac{\overline{w}-\overline{w}_0\overline{\rho}_0}{\overline{w}_0\overline{\rho}_0 - \rho_{21}\overline{w}}. \tag{27}$$

The correlation (27) shows that analogs of the interaction coefficients from the Kolmogorov theorem contain the common part $\overline{w}_0\overline{\rho}_0$ with the opposite sign (like in the third Newton's law concerning the two acting bodies) and similar terms $\overline{w}$, $\rho_{21}\overline{w}$, where density ratio in from of one of them underlines influence of the density ration on the interaction of the phases in a two-phase flow. It looks reasonable from the physical point of view, therefore solution (7) may be used for analysis in non-stationary case too, with the coefficients (27).

As far as probability is a value between 0 and 1, from the above follows that always $\gamma > 0$, which yields from the above the following conditions satisfying this requirement:

$$\overline{w}_0\overline{\rho}_0 < \overline{w} < \frac{\overline{w}_0\overline{\rho}_0}{\rho_{21}}, \quad \rho_{21} < 1; \quad \text{or} \quad \frac{\overline{w}_0\overline{\rho}_0}{\rho_{21}} < \overline{w} < \overline{w}_0\overline{\rho}_0, \quad \rho_{21} > 1.$$

The last one shows that two-phase flow cannot be arbitrary from the considered expressions. If density of the phase 2 is less than density of phase 1, the velocity of two-phase mixture is variated in a range, which is expanding with decrease of the density of a second phase. Similarly, in case of denser second phase, the flow velocity is supposed to be "allowed" for change in a specified range from $\bar{w}_0 \bar{\rho}_0$ to a value aiming to zero with increase of density ratio.

5. **The conclusions**

Analysis of the derived equation array has shown the correctness of the assumptions made and the basic conservation equations. The conditions for correctness of the model in a form of some limitations on the correlation between density ratio of the phases and velocity of the two-phase flow were obtained from analysis performed. The results obtained need comparison with the results of computations available from the other authors in a literature, and, the most important, with the experimental data. This is a subject for further investigation.

**REFERENCES**

1. Balescu Radu. Statistical mechanics of charged particles. - New York: Interscience Publishers. - 1963. - 465 pp.
2. Bogoliubov N.N. Problems of Dynamic Theory in Statistical Physics.- Oak Ridge, Tenn.: Technical Information Service.- 120pp.
3. Davis S.H. Contact-Line Problems in Fluid Mechanics// J. Appl. Mech.- 1983.- V. 50 (4b).- P. 977-982.
4. Zhezherin R.P. "The "electromagnetic crucible" problem," in: Problems of Magneto-hydrodynamics and Plasma Dynamics, Izd. AN Latv. SSR, Riga (1959), P. 279–294 (in Russian).
5. Ostroumov G.A. Interaction of electric and hydrodynamic fields.-M.: Nauka.- 1979.- 320 pp. (in Russian).
6. Ashby W.R. An Introduction to Cybernetics.- London: Chapman & Hall.- 1957.- 294 pp.
7. Benoit B. Mandelbrot. The Fractal Geometry of Nature.- San Francisco: W.H. Freeman, 1983.- 480 p.
8. Samko, S.; Kilbas, A.A.; and Marichev, O. Fractional Integrals and Derivatives: Theory and Applications.- Publisher: Taylor & Francis Books, 1993. - 1006 p.
9. Babenko, Yu.I. Heat-Mass Transfer: Methods for Calculation of Thermal and Diffusional Fluxes.- Leningrad: Khimia Publ, in Russian, 1986.- 236 p.
10. Kolmogorov A.N. The general theory of dynamical systems and classical mechanics. Proceedings of the International Congress of Mathematicians (Amsterdam, 1954), Vol. 1, pages 315-333.
11. Wentzel E.S. Operations research. - M .: Publishing house "Soviet radio", 1972 - 407 p.
12. Nakorchevski A.I. Heterogeneous turbulent jets. Kyiv: Naukova Dumka, 1980, 142 p.
13. Nakorchevski A.I., Kazachkov I.V. Calculation of the heterogeneous turbulent jet. In book: Systems of automation of continuous technological processes, Institute of Cybernetics of NASU, 1979, P. 68-79.
14. Kazachkov I.V. Der turbulente versenkte Strahl von zweier unmischbaren Fluessigkeiten. Beitrag. Inst. fuer Kybernetik der Akademie der Wissenschaften der Ukraine. Kiev, 1980. 20 p.
15. Kazachkov I.V. Mathematical modeling of heterogeneous turbulent jets in cylindrical chamber// Soviet automatic control, 1980, vol.13, jan.-feb., p. 1-6.
16. Kazachkov I.V., Nakorchevski A.I. The stream lines in a turbulent two-phase jet of two immiscible liquids/ Abstracts of the V All SU Meeting on Theor. And Appl. Mechanics. Alma-Ata, 1981 (In Russian).
17. Kazachkov I.V. Investigation of a turbulent mixing and wall protecting by garnissage in the jet devices working on two immiscible fluids. Candidate of Physics and Mathematics Dissertation (Ph.D.). Kiev T.G. Shevchenko State Univ. 1981. 138 p. (in Russian).
18. Nakorchevski A.I., Basok B.I. Hydrodynamics and heat transfer in heterogeneous systems and devices of pulsating type. Kyiv: Naukova Dumka, 2001, 348 p.